\documentclass[twocolumn]{aastex7}

\begin{document}

\title{Redshift Distributions of Fast Radio Bursts Inferred Using Clustering in Dispersion Measure Space}

\author[orcid=0000-0003-3929-3229, gname=Hui, sname=Peng]{Hui Peng}
\affiliation{Department of Astronomy, School of Physics and Astronomy, Shanghai Jiao Tong University,
Shanghai 200240, People’s Republic of China}
\affiliation{Key Laboratory for Particle Astrophysics and Cosmology (MOE)/Shanghai Key Laboratory for Particle Physics and Cosmology, Shanghai 200240, People’s Republic of China}
\email{pengh.me@gmail.com}

\author[orcid=0000-0002-9359-7170, gname=Yu, sname=Yu]{Yu Yu}
\affiliation{Department of Astronomy, School of Physics and Astronomy, Shanghai Jiao Tong University,
Shanghai 200240, People’s Republic of China}
\affiliation{Key Laboratory for Particle Astrophysics and Cosmology (MOE)/Shanghai Key Laboratory for Particle Physics and Cosmology, Shanghai 200240, People’s Republic of China}
\email[show]{yuyu22@sjtu.edu.cn}

\correspondingauthor{Yu Yu}

\begin{abstract}
Fast radio bursts (FRBs), millisecond-duration radio transient events, possess the potential to serve as excellent cosmological probes.
The FRB redshift distribution contains information about the FRB sources, providing key constraints on the types of engines.
However, it is quite challenging to obtain the FRB redshifts due to the poor localization and the faintness of the host galaxies.
This reality severely restricts the application prospects and study of the physical origins of FRBs.
We propose that the clustering of observed FRBs can be an effective approach to address this issue without needing to accurately model dispersion measure (DM) contributions from the host galaxy and the immediate environment of the source.
Using the clustering of $5\times 10^7$ simulated FRBs from future observations with sensitivity similar to the second phase of the Square Kilometre Array, we show that in extragalactic DM space, the redshift distributions can be accurately reconstructed, and the mean redshift for FRBs between 384.8 and 1450.3~$\rm pc\,cm^{-3}$ can be constrained to $\sim\!0.001\pm0.003 (1+z)$.
The results demonstrate the potential of FRB clustering to constrain redshift distributions and provide valuable insights into FRB source models and cosmological applications.
\end{abstract}



\section{\label{sec:into}Introduction}
Fast radio bursts (FRBs) are millisecond-duration transients in the radio band predominantly from cosmological distances~\citep{Petroff:2019aa}.
Since the first discovery in 2007 ~\citep{Lorimer:2007aa}, hundreds of FRBs have been observed in recent years by many radio telescopes, such as Parkes~\citep{Thornton:2013aa}, the Canadian Hydrogen Intensity Mapping Experiment \citep[CHIME;][]{CHIME/FRB-Collaboration:2021aa,CHIME-Collaboration:2022aa}, the Australian Square Kilometre Array Pathﬁnder \citep[ASKAP;][]{Bannister:2017aa,Shannon:2018aa}, and the Five-hundred-meter Aperture Spherical Radio Telescope \citep[FAST;][]{Luo:2020aa,Niu:2022aa}.
With rapid growth of detected events, FRBs have great potential to serve as powerful cosmological probes to trace the ionized gas in the intergalactic medium \citep[IGM;][]{Macquart:2020aa}, provide information about the galaxy environments \citep{Niu:2022aa,Jow:2024aa} and study the large-scale structure of Universe \citep{Rafiei-Ravandi:2021aa,Lee:2022aa}.

The physical origin and mechanism of FRBs still remain unsolved puzzles to date \citep{Platts:2019aa,Xiao:2021aa,Zhang:2023aa}.
Studies suggest that at least some FRBs are produced by magnetars \citep[e.g.,][]{Bochenek:2020aa,CHIME/FRB-Collaboration:2020aa,Lu:2020aa,Yu:2021aa,CHIME/FRB-Collaboration:2022aa,Wang:2022aa,Bruni:2024aa,Hu:2024aa}.
Different source models propose different FRB redshift distributions, such as tracking the star formation history or binary neutron star merger events (a significant delay with respect to star formation) \citep[e.g.,][]{Munoz:2016aa,Macquart:2018aa,Margalit:2019aa,Wang:2020aa,Zhang:2021ab,James:2022ac,Zhang:2022aa,Qiang:2022aa,Sharma:2024aa}.
Therefore, to figure out the engines that power FRBs, it is crucial to accurately determinate their intrinsic redshift distributions.
Meanwhile, the research into the FRB population is also important for harnessing them as cosmological probes.
The cosmological studies using FRBs make the best of the dispersion measure (DM), which integrates the column density of ionized electrons from the source to Earth.
The DM can be a rough proxy of redshift, based on the theoretically motivated relation between the DM and redshift.
The DM-$z$ relation is also known as the “Macquart relation,” owing to the fact that it is observationally verified with localized FRBs in \cite{Macquart:2020aa}.
A statistically significant cross correlation has been found between FRB DMs and galaxies in the photometric galaxy surveys \citep{Rafiei-Ravandi:2021aa}.
With redshifts of large amounts of FRB host galaxies obtained, FRBs can provide precise constraints on the dark energy, baryon density, the Hubble constant and other cosmological studies \citep[e.g.,][]{James:2022ab,Petroff:2022aa,Zhang:2023ab,Saga:2024aa,Wei:2024aa}.

However, it is less feasible to accurately measure the redshifts of many FRB hosts.
In order to associate FRBs with their host galaxies, the arcsecond localization is fundamentally required and must be restricted to within $\sim\! 0.1^{\prime \prime}-0.5^{\prime \prime}$ for $z>1$ events \citep{Macquart:2015aa}.
Due to the poor resolution of the radio telescopes (e.g., sub-arcminute precision of CHIME \citep{Michilli:2021aa,CHIME/FRB-Collaboration:2024aa}), it is difficult to localize FRBs sufficiently well to identify and measure the redshifts of the host galaxies.
At the same time, obviously different from the specific bright targets in galaxy redshift surveys like the Dark Energy Spectroscopic Instrument \citep[DESI;][]{DESI-Collaboration:2022aa}, FRB host galaxies could be intrinsically faint \citep{Marnoch:2023aa} and some of them are localized at low Galactic latitude (e.g., ASKAP, FAST), which makes it a tough task to obtain their spectroscopic redshifts \citep{Glowacki:2024aa}.
Therefore, even for the upcoming high-sensitivity radio telescope with sub-arcsecond localization precision, Square Kilometre Array \citep[SKA;][]{Macquart:2015aa}, obtaining the accurate redshifts of FRBs is still full of challenges.
So far, only dozens of FRBs are localized with their host galaxies and measured redshifts, around 10\% of all the detected FRBs.
For the unlocalized FRBs, the pseudo-redshifts estimated using simulations and modeling DM contributions from the host galaxy and the local environment of the source \citep[e.g.,][]{Zhang:2021aa} will exacerbate the systematic error in analysis due to the degeneracy between different components of observed DMs \citep{James:2022aa}.
With the rapid increase in detected FRBs from future radio telescopes, it is urgent to find a method independent of spectroscopic measurements to accurately derive the redshift distribution of unlocalized FRBs in order to fully leverage this data.

In addition to the direct spectroscopic redshift measurements, the clustering of FRBs has the potential to address this challenge \citep[e.g.,][]{Shirasaki:2017aa,Rafiei-Ravandi:2020aa}.
Compared to locating FRB host galaxies, large-scale structure is less sensitive to localization precision and can be effectively utilized in various future FRB observations.
With a large sample size of detected FRBs in future observations, obtaining and utilizing the FRB clustering signals in DM space is very promising for inferring redshift distributions.
Instead of cross-correlating galaxy samples, we propose using a verified redshift self-calibration method in galaxy redshift surveys to directly infer redshift distributions from observed FRB DMs and angular positions \citep[e.g.,][]{Peng:2022aa,Xu:2023aa,Peng:2023aa,Peng:2024aa,Zheng:2024aa}.
The basic idea is that the observed correlation signals between different tomographic bins stem from the combination of galaxy auto correlations in redshift bins \citep{Schneider:2006ta,Zhang:2010wr,Zhang:2017um,Schaan:2020up}.
Then the true redshift distributions can be effectively inferred from these non-zero galaxy-galaxy correlations.
Similarly, since the DM can be regarded as rough proxy of redshift, we can make use of the clustering in DM space to infer the redshift distributions of FRBs.

In this work, we demonstrate that the redshift distributions of FRBs can be accurately reconstructed in DM space by utilizing clustering from future observations.
The simulated FRBs we use are presented in Section~\ref{sec:data}.
Section~\ref{sec:method} gives a brief overview of the method and the algorithm.
The results are presented in Section~\ref{sec:results}.
We conclude and discuss in Section~\ref{sec:dis_cons}.

\section{\label{sec:data}Mock Data Sets}
\subsection{Uncertainties in DM}
The DM represents the free electron density integrated along the line of sight to the FRB,  ${\rm DM}=\int n_e/(1+z)dl$, where $l$ is the proper line element. 
The observed DM of an FRB can be divided into several components:
\begin{equation}
    \mathrm{DM}_{\mathrm{obs}}=\mathrm{DM}_{\mathrm{MW}}+\mathrm{DM}_{\mathrm{E}}\ ,
    \label{DM_obs}
\end{equation}
where $\mathrm{DM}_{\mathrm{MW}}$ is the contribution from our Galactic interstellar medium ($\mathrm{DM}_{\mathrm{MW,\,ISM}}$) and our Galactic halo ($\mathrm{DM}_{\mathrm{MW,\,halo}}$), and $\mathrm{DM}_{\mathrm{E}}$ is the extragalactic contribution.
The $\rm DM_{MW,\,ISM}$ can be derived from the typical models like NE2001 \citep{Cordes:2002aa} or YMW16 \citep{Yao:2017aa}, and $\rm DM_{MW,\,halo}$ can be properly estimated based on studies \citep[e.g.,][]{Prochaska:2019aa,Keating:2020aa,Platts:2020aa,Yamasaki:2020aa}.
The extragalactic component can be separated as
\begin{equation}
    \mathrm{DM}_{\mathrm{E}}=\mathrm{DM}_{\mathrm{IGM}}+\frac{\mathrm{DM}_{\text {host }}}{1+z}\ , 
    \label{DM_E}
\end{equation}
where $\mathrm{DM}_{\mathrm{IGM}}$ is the dominant contribution from the diffuse IGM gas that traces the large-scale structure of Universe, and $\mathrm{DM}_{\text {host}}$ is from the host galaxy and the progenitor's immediate environment.
The IGM component of DM is a function of redshift and can be derived theoretically by \citep{Ioka:2003aa,Inoue:2004aa,Deng:2014aa}
\begin{equation}
    \left\langle\mathrm{DM}_{\mathrm{IGM}}(z)\right\rangle=\frac{3 c H_0 \Omega_b f_{\mathrm{IGM}}}{8 \pi G m_p} \int_0^z \frac{\chi\left(z^{\prime}\right)\left(1+z^{\prime}\right)}{E\left(z^{\prime}\right)}d z^{\prime},
    \label{DM_IGM}
\end{equation}
where $G$ is the gravitational constant, $m_p$ is proton mass; $\Omega_b$ is the energy density fraction of baryons; $f_{\mathrm{IGM}}$ is the fraction of baryons in the IGM and is adopted as 0.83; $E(z)\equiv H(z)/H_0$ is the dimensionless Hubble parameter; and $\chi(z) \simeq \frac{3}{4} \chi_{\mathrm{e}, \mathrm{H}}(z)+\frac{1}{8} \chi_{\mathrm{e}, \mathrm{He}}(z)$, where $\chi_{\mathrm{e}, \mathrm{H}}$ and $\chi_{\mathrm{e}, \mathrm{He}}$ are the fractions of ionized electrons in hydrogen and helium, respectively.
For redshifts $z\leqslant3$, the intergalactic hydrogen and helium are both fully ionized \citep{Fan:2006aa}, and thus $\chi(z)\simeq 7/8$.

Note that Equation~(\ref{DM_IGM}) gives average values; the measured DM for individual FRBs will deviate from the theoretical value caused by the plasma density fluctuations due to large-scale structures.
For an FRB with $\rm DM_{obs}$, $\rm DM_{MW}$ determined, the uncertainty of $\rm DM_E$ can be given by
\begin{equation}
    \sigma_{\mathrm{E}}=\sqrt{\sigma_{\mathrm{obs}}^2+\sigma_{\mathrm{MW}}^2+\sigma_{\mathrm{IGM}}^2+\left(\frac{\sigma_{\mathrm{host}}}{1+z}\right)^2}\ .
    \label{uncertainty}
\end{equation}
Here $\sigma_{\mathrm{obs}}$, $\sigma_{\mathrm{MW}}$, $\sigma_{\mathrm{IGM}}$ and $\sigma_{\mathrm{host}}$ are uncertainties from observation and different DM components, $\rm DM_{MW}$, $\rm DM_{IGM}$, and $\rm DM_{host}$ respectively.
Following \cite{Zhang:2023ab}, we can adopt $\sigma_{\mathrm{obs}}=0.5\, \mathrm{pc}\, \mathrm{cm}^{-3}$ \citep{CHIME/FRB-Collaboration:2021aa} and $\sigma_{\mathrm{MW}}=10\, \mathrm{pc}\, \mathrm{cm}^{-3}$ \citep{Manchester:2005aa}.
The deviation from the central values of $\rm DM_{IGM}$ and $\rm DM_{host}$ is characterized by $\sigma_{\mathrm{IGM}}(z)=173.8\,z^{\mathrm{0.4}}\, \mathrm{pc}\, \mathrm{cm}^{-3}$ \citep{Qiang:2020aa} and $\sigma_{\mathrm{host}}=30\, \mathrm{pc}\, \mathrm{cm}^{-3}$\citep{Zhang:2023ab}, respectively.

\subsection{Redshift Distribution and Event Rate}
To date, due to the deficiency in localized FRBs, the intrinsic FRB redshift distribution remains unclear.
In the meantime, the detected distribution is closely related to the telescope’s sensitivity threshold and instrumental selection effects, making it difficult to characterize in reality \citep{James:2022ac,Qiang:2022aa,Zhang:2022aa}.
Here we assume that FRBs have a constant comoving number density \citep{Munoz:2016aa}.
Consequently, the number of FRBs within a shell of thickness $dz$ at redshift $z$ is proportional to the shell’s comoving volume, $[4\pi d_C^2(z)/ H(z)]~dz$, divided by $(1+z)$ to reflect the impact of cosmological time dilation on the burst rate.
Then a Gaussian cutoff at some redshift $z_{\text {cut }}$ is introduced to represent an instrumental signal-to-noise threshold, and the redshift distribution function reads
\begin{equation}
    P_{\text {const }}(z) \propto \frac{d_L^2(z)}{(1+z)^3 H(z)} \exp \left(-\frac{d_L^2(z)}{2 d_L^2\left(z_{\text {cut }}\right)}\right)\ ,
    \label{nz}
\end{equation}
where $d_L(z)=d_C(z)(1+z)$ is the luminosity distance.
The Gaussian cutoff $z_{\text {cut }}= 1.0$ is adopted \citep{Li:2018aa,Qiang:2021aa,Zhang:2023ab}.
Note that while different $z_{\text {cut }}$ values will alter the FRB redshift distribution, they primarily affect the relative reconstruction accuracy in different DM spaces rather than the overall validity of the method.
Practically, we might empirically adjust the redshift distribution through FRB DM selection.

We consider a future specialized FRB survey where the event rate will be generated based on the SKA survey, which is expected to operate for 50 yr and more \citep{Torchinsky:2016aa}.
For the mid-frequency aperture array of the phase 1 (SKA1-MID), the estimated all-sky event rate is $N_{\rm sky}\sim\! 3\times 10^6~\rm {sky}^{-1}~{day}^{-1}$ on the basis of the results from Parkes and ASKAP FRB surveys \citep{Zhang:2023ab}.
The value will increase to $\sim\!9\times10^7~\rm {sky}^{-1}~{day}^{-1}$ for the second phase of SKA (SKA2-MID) \citep{Wei:2024aa}.
Then, for the exposure time $t_{\rm obs}$, the estimated number of detected FRBs can be given by
\begin{equation}
    N_{\text {exp }}= t_{\rm obs}\Omega N_{\text {sky }}\ ,
    \label{N_exp}
\end{equation}
where $\Omega$ is the sky coverage fraction of the instantaneous observation, as FRBs are transient events.
Here we take the effective instantaneous field of view of SKA2-MID as 200 $\rm deg^2$ \citep{Torchinsky:2016aa,Hashimoto:2020aa} and utilize it to calculate the $\Omega$.
For the $t_{\rm obs}$ on SKA2-MID FRB search, an average of 20\% of observing time per year is assumed, and $t_{\rm obs} = 20\% \times 365 ~\rm day~yr^{-1}$ \citep{Bhandari:2019aa,Zhang:2023ab}.
Therefore, the event rate of FRBs by SKA2-MID can be $\sim\! 3\times 10^7 ~\rm yr^{-1}$.

For a specific FRB survey with configurations similar to SKA2-MID, we have a much larger value of $t_{\rm obs}$ and thus can detect more FRBs.
Instead of observing $\sim\! 30000~\rm deg^2$ for SKA2-MID \citep{Torchinsky:2016aa}, here we consider a relatively limited sky coverage during the 1 yr observation to obtain high number density.
In our mock datasets, we adopt the values of $5\times 10^7 $ and $1000~\rm deg^2$ as the fiducial numbers of detected FRBs and sky coverage to estimate the statistical uncertainty of a 1 yr observation, respectively.
These utilized parameters are based on SKA, but the analysis generalizes to all future FRB observations with similar sensitivity and sky coverage, particularly those with poor localization resolution.

\subsection{Mock FRB Data and Statistics}
\begin{figure}
\centering
	\includegraphics[width=8cm]{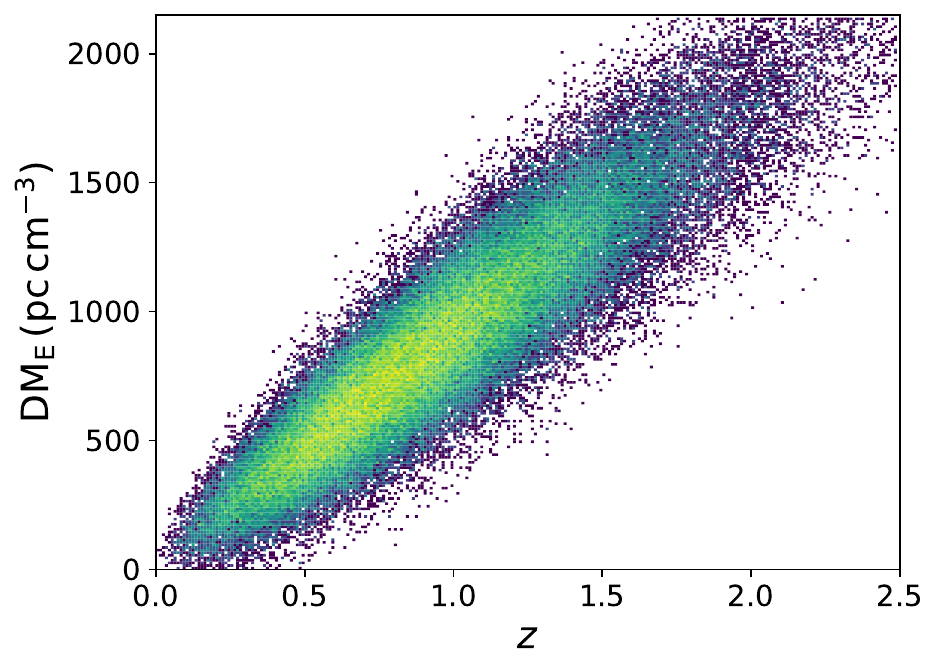}
    \caption{The comparison between the extragalactic DMs and redshifts of mock FRB data. The upper limits of $\rm DM_E$ and $z$ are set to $\rm 2150~pc\,cm^{-3}$ and 2.5, respectively.}
    \label{fig:DM_scatter}
\end{figure}

\begin{figure*}
\centering
	\begin{minipage}{7.5cm}
    	\centering
        \includegraphics[width=7.5cm]{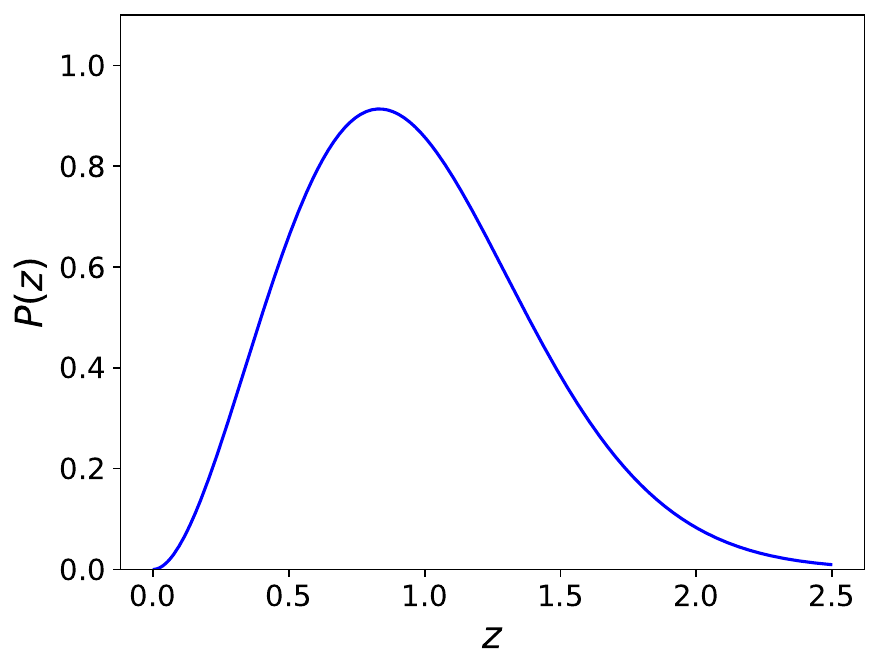}
    \end{minipage}
    \hspace{0.2cm}
	\begin{minipage}{7.5cm}
    	\centering
        \includegraphics[width=7.5cm]{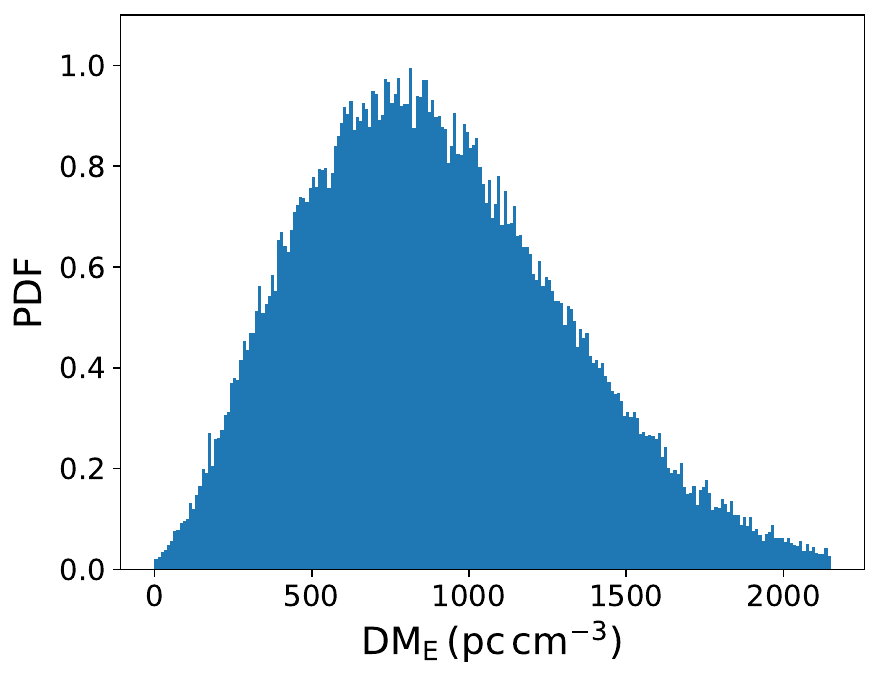}
    \end{minipage}
    \caption{Left panel: the redshift distribution adopted to simulate FRBs ranging from $z=0$ to $z=2.5$. Right panel: the extragalactic DM distribution ($\rm 0~pc\,cm^{-3}< DM_E<2150~pc\,cm^{-3}$) of mock FRB data.}
    \label{fig:mock_frbs}
\end{figure*}
We construct the mock datasets in the following steps.
Firstly, we derive the relation between $\rm DM_{IGM}$ and redshifts in Equation~(\ref{DM_IGM}) by adopting a fiducial $\Lambda$CDM cosmology.
The dimensionless Hubble parameter, $E(z)$, is given by
\begin{equation}
    E(z)=\sqrt{\Omega_m(1+z)^3+\Omega_\Lambda}\ ,
\end{equation}
where $\Omega_m$ is the present fractional density of matter, and vacuum energy density $\Omega_\Lambda=1-\Omega_m$.
Then, we focus on the $\rm DM_E$, as it can be obtained by subtracting the well-understood $\rm DM_{MW,ISM}$ and $\rm DM_{MW,halo}$ from the measured $\rm DM_{obs}$ in practical observations.
Using theoretical values of $\rm DM_{IGM}$, we assume $\rm DM_{\mathrm{host}}=50\, \mathrm{pc}\, \mathrm{cm}^{-3}$ to calculate the average values of $\rm DM_E$ for the simulated FRBs.
Finally, according to Equation~(\ref{nz}), we generate the FRB redshifts and obtain their observed $\rm DM_E$ considering a Gaussian uncertainty presented in Equation~(\ref{uncertainty}).
In Figure~\ref{fig:DM_scatter}, we show the comparison between the redshifts and generated $\rm DM_E$ of simulated FRBs.
The $\rm DM_E$ range of mock FRB data is constrained within $\rm 0< DM_E< 2150~pc\,cm^{-3}$, and the upper limit of redshifts is set to 2.5.
In practice, a tiny fraction of detected FRBs within this $\rm DM_E$ range may fall outside the stated edge of redshifts, but this is unlikely to have a significant impact on the analysis or the key conclusions.
The distributions of mock FRB redshifts and their $\rm DM_E$ are presented in Figure~\ref{fig:mock_frbs}.

From the redshift distributions of FRBs in $\rm DM_E$ bins, we can calculate the clustering theoretically.
The angular power spectrum between two DM bins $i$ and $j$ can be derived by applying the Limber approximation, i.e.,
\begin{equation}
C_{ij}^{gg}(\ell)= \int_0^{\chi_{\mathrm{H}}} \frac{d \chi}{\chi^2} K_i(\chi) K_j(\chi) P_{\delta}\left(k=\frac{\ell+1 / 2}{\chi}, z\right)\ ,
\end{equation}
with $K(\chi)= n(z) b_g(z) H(z)$.
$\chi_{\mathrm{H}}$ and $\chi$ are the comoving horizon distance and the comoving radial distance, respectively. $n(z)$ is the redshift distribution within DM bin, $b_g(z)$ is the linear galaxy bias, and $H(z)=dz/d\chi$ is the expansion rate.
$P_{\delta}$ is the matter power spectrum.
Here, we assume an ideal linear host galaxy bias follows $b_g(z)=1.5/D(z)$, where $D(z)$ is the linear growth factor normalized by $D(z = 0)=1$.
Throughout the work, we adopt a fiducial $\Lambda$CDM cosmology with parameter values from the Planck 2018 TT,TE,EE+lowE+lensing+BAO results \citep{Planck-Collaboration:2020aa}.


\section{\label{sec:method}Method}
We utilize the self-calibration method, which attempts to reconstruct the true redshift distributions from the galaxy clustering of observed data itself \citep{Schneider:2006ta,Zhang:2010wr}.
Assume that we split detected FRBs into $n$ DM bins.
We denote the ratio of the FRBs in redshift bin $i$ that are observed in DM bin $j$ as $P_{ij}\,{\equiv}\,N_{i{\rightarrow}j}/N_j^D$.
Here, $N_{i{\rightarrow}j}$ represents the number of FRBs originating from redshift bin $i$ and detected within DM bin $j$, and $N_j^D$ is the total number of FRBs within DM bin $j$.
For each DM bin $j$, there exists the normalization $\sum_iP_{ij}=1$.
Note that for each DM bin $j$, the $P_{ij}$ scatter is equivalent in meaning to the redshift distribution from redshift bins, and thinner bins result in a more detailed distribution.

The angular power spectrum of two FRB DM bins $C_{ij}^{gg,D}$ and redshift bins $C_{ij}^{gg,R}$ can be connected through the following relationship:
\begin{equation}
    C_{ij}^{gg,D}(\ell)=\sum_kP_{ki}P_{kj}C_{kk}^{gg,R}(\ell)+\delta{N_{ij}^{gg,D}(\ell)}\ .
    \label{eqn:CggDsum}
\end{equation}
The last term $\delta{N_{ij}^{gg,D}}$ is the associated shot noise fluctuation after the subtraction of the ensemble average, where the fluctuation level can be well approximated by a Gaussian distribution with
\begin{equation}
    \sigma_{i j}^{g g, D}=\sqrt{\frac{1}{(2 \ell+1) \Delta \ell f_{\text{sky}}} \frac{1+\delta_{i j}}{\bar{n}_{i} \bar{n}_{j}}}\ , 
    \label{eqn:noise}
\end{equation}
where $\Delta\ell$ is the $\ell$ bin size, $f_\mathrm{sky}$ is the sky fraction, $\bar{n}_i$ ($\bar{n}_j$) is the number density of FRBs in DM bin $i$ ($j$), and $\delta_{i j}$ is the Kronecker delta.
For a given $\ell$, We can rewrite the Equation~(\ref{eqn:CggDsum}) in matrix form,
\begin{equation}
    C_{\ell}^{gg,D}=P^{T}C_{\ell}^{gg,R}P+\delta{N_{\ell}^{gg,D}}\ .
    \label{eqn:CD}
\end{equation}
We denote the matrix $P$ here as the scattering matrix.
Therefore, by solving the Equation~(\ref{eqn:CD}), we can use the correlations of observed FRBs on the left side to reconstruct $P$ and thus obtain the true redshift distributions in each DM bin.
The adopted algorithm is same as the one in \cite{Peng:2024aa}, which can reconstruct accurate results through the minimization of objective function
\begin{equation}
    \mathcal{J} = \frac{1}{2} \sum_{\ell}\left\|V_{\ell}^{1 / 2} \circ \left(C_{\ell}^{g g, D}-P^{T} C_{\ell}^{g g, R} P\right)\right\|_{F}^{2}\ ,
    \label{define:J}
\end{equation}
where $\|.\|_F$ is the Frobenius form, the open circle dot operator $\circ$ indicates the Hadamard (element-wise) product, and the weight $V_{\ell}$ is the inverse variance matrix of each matrix element of $C_{\ell}^{g g, D}$.
The self-calibration algorithm utilizes novel non-negative matrix factorization (NMF) technique and has shown extraordinary reconstruction ability in the applications to the samples from galaxy redshift surveys.
We encourage the readers to consult the referenced paper for the details.

We calculate the angular power spectra with \texttt{CCL}\footnote{\href{https://github.com/LSSTDESC/CCL}{https://github.com/LSSTDESC/CCL}} \citep{Chisari:2019wv} and utilize $\ell$ modes within $100\leqslant\ell<1000$ which is divided into six broad bands with edges $\ell=$ 100, 418, 583, 711, 819, 914, and 1000.
We generate 30 mock power spectrum observations, each containing the theoretical power spectrum described above and the corresponding Gaussian random noise generated according to Equation~(\ref{eqn:noise}).
When we apply the algorithm to the input $C_{\ell}^{gg,D}$, Gaussian perturbations are applied to the data once again to prevent the solution from converging to local minima.
The initial guess strategy and the selection criterion for the results are the same as in \cite{Peng:2024aa}.
Each mock dataset is perturbed 100 times.
We take the median values of the reconstructed scattering matrix elements under the perturbations as the best fit for each mock dataset, with renormalization to 1 in each column.

\section{\label{sec:results}Results}
\begin{figure}
\centering
	\includegraphics[width=8cm]{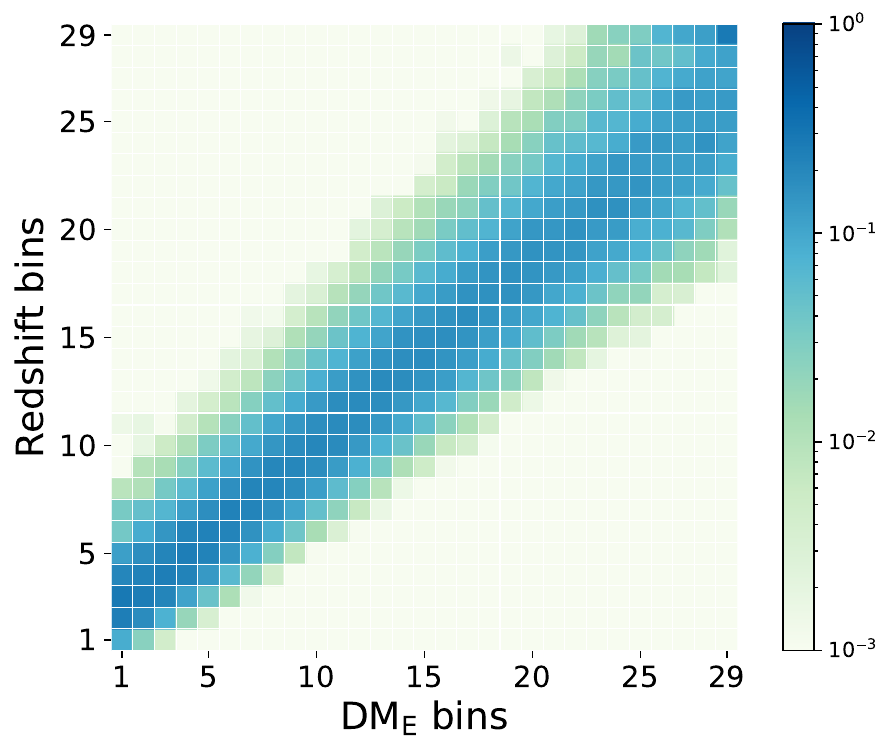}
    \caption{Visual representation of the scattering matrix $P$ in Equation~(\ref{eqn:CD}) for mock FRB data from the adopted binning scheme.
    Each element $P_{ij}$ represents the fraction of the FRBs in redshift bin $i$ that are detected in $\rm DM_E$ bin $j$.
    For each column, the values of elements sum to 1.}
    \label{fig:P_true}
\end{figure}

\begin{figure*}
    \centering
	\includegraphics[width=15.5cm]{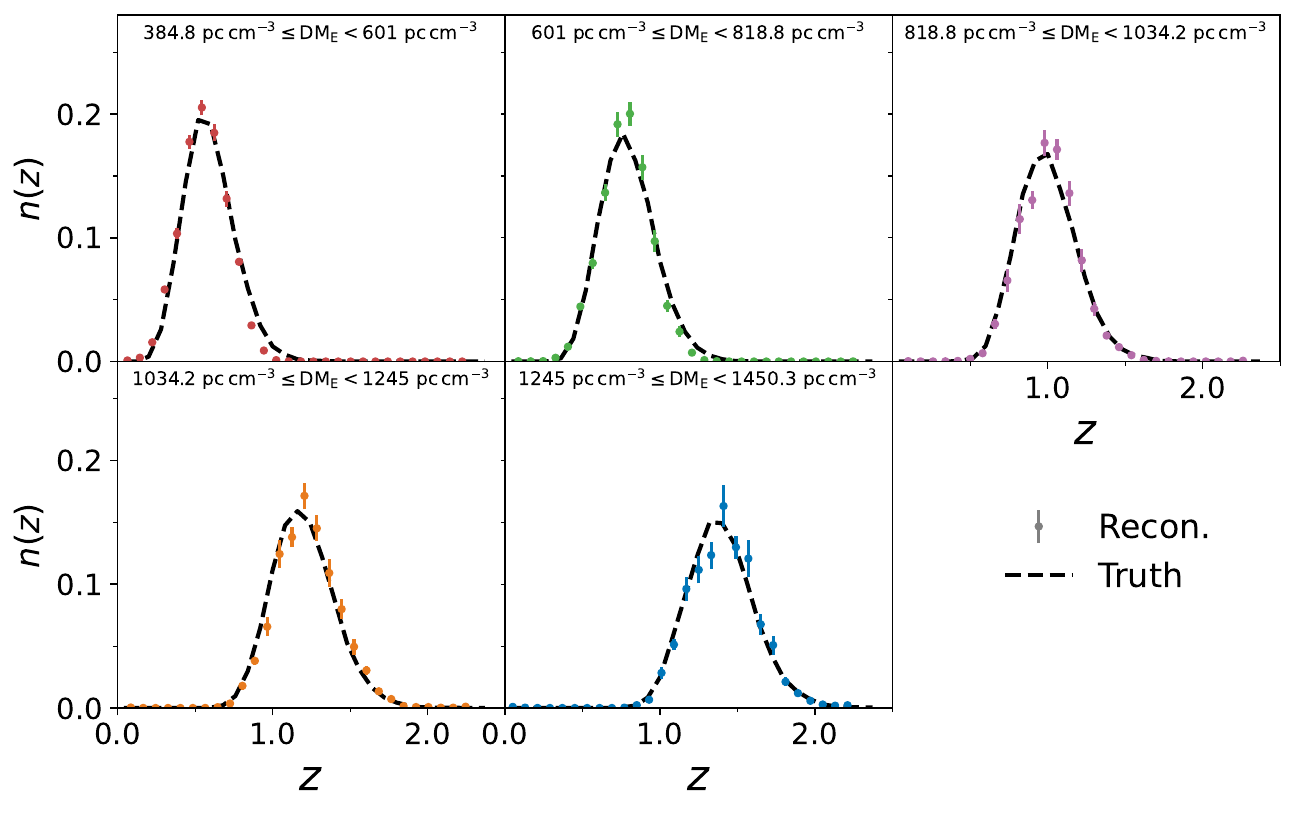}
    \caption{The reconstructed redshift distributions for mock FRB data in five DM bins, using $5\times 10^7 $ FRBs within $1000~\rm deg^2$ sky coverage. The points and the bars represent the mean values and 1$\sigma$ uncertainties among 30 mock runs. The true redshift distributions are shown in black dashed lines for comparison.}
    \label{fig:nz_recon}
\end{figure*}

\begin{figure}
\centering
	\includegraphics[width=8cm]{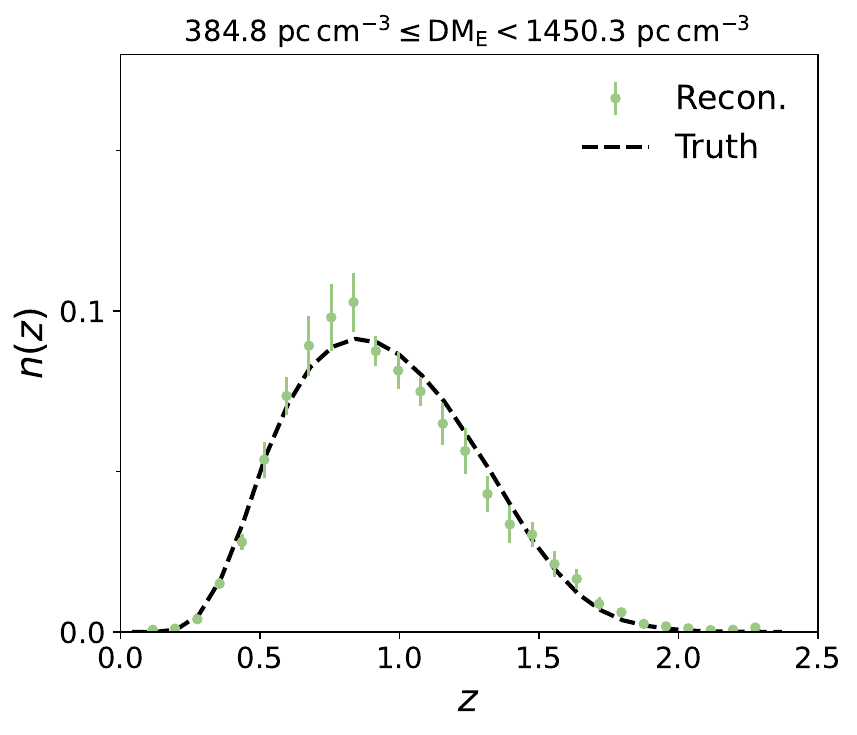}
    \caption{Similar to Figure~\ref{fig:nz_recon}, but for all mock FRB data between 384.8 and 1450.3~$\rm pc\,cm^{-3}$.}
    \label{fig:nz_recon_sample0}
\end{figure}

\begin{figure*}
    \centering
	\includegraphics[width=15.5cm]{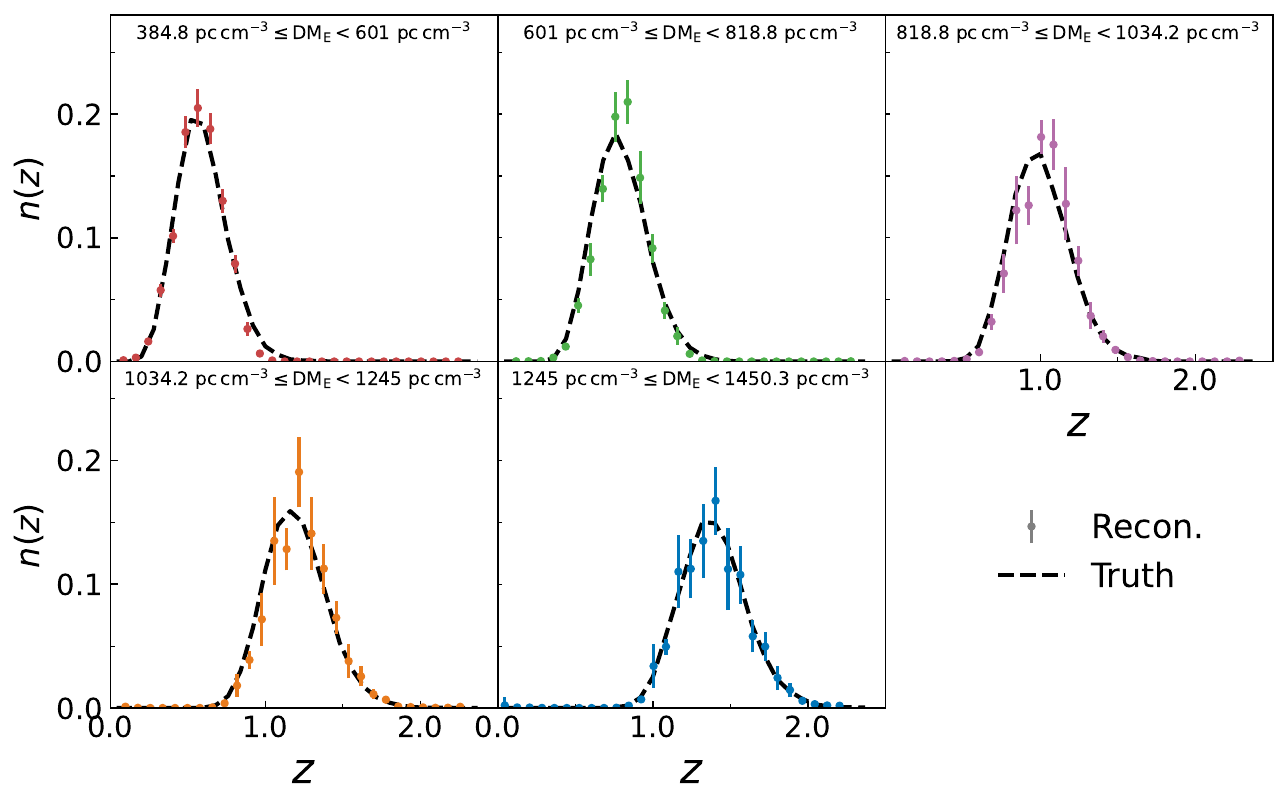}
    \caption{Same as Figure~\ref{fig:nz_recon}, but the sky coverage is increased from 1000~$\rm deg^2$ to 2000~$\rm deg^2$, while the number of detected FRBs remains unchanged.}
    \label{fig:nz_recon_compare}
\end{figure*}

Before dividing the mock FRB data into several $\rm DM_E$ bins, we can utilize the $\rm DM_{E}$ to roughly estimate the corresponding redshift $z_{\rm DM}$ directly based on the ${\rm DM_{IGM}}-z$ relation in Equation~(\ref{DM_IGM}).
The $z_{\rm DM}$ is only used to roughly determine the binning scheme and does not affect the analysis.
In contrast to the photometric redshift errors (typically $\sigma_z \leqslant 0.05(1+z)$), there is a much larger scatter between $z_{\rm DM}$ and the true redshifts, which can be roughly approximated by $\sigma_z \sim\! 0.1(1+z)$.
This large scatter in the data makes inferring the redshift distributions from the self-calibration method more challenging and increases uncertainty in the reconstruction.
Using the estimated $z_{\rm DM}$, we can select the binning scheme and then obtain the $\rm DM_E$ edges of each bin.
The edges of the redshift bins are the same as the $z_{\rm DM}$ values adopted to determine the $\rm DM_E$ binning scheme.
In practice, we can also assign required values of $\rm DM_E$ as specific edges of the bins in the binning scheme.
In this way, we do not need to accurately model $\rm DM_{host}$ when reconstructing the redshift distributions; instead, we regard it as part of the uncertainty of $\rm DM_{IGM}$.

Based on the estimated $z_{\rm DM}$ we can divide the mock FRB data into 28 bins of equal width in the range $0 < z_{\rm DM} < 2.24$ with bin widths set to 0.08 and one additional wide bin with $2.24 \leqslant z_{\rm DM} < 2.5$.
With this binning scheme, the true scattering matrix $P$ in Equation~(\ref{eqn:CD}) is shown in Figure~\ref{fig:P_true}.
In order to demonstrate the reconstruction performance, we consider FRBs within $0.4 \leqslant z_{\rm DM} < 1.6$ and divide the data into five equal-width sequential $z_{\rm DM}$ bins.
The corresponding $\rm DM_E$ edges of five bins are 384.8~$\rm pc\,cm^{-3}$, 601~$\rm pc\,cm^{-3}$, 818.8~$\rm pc\,cm^{-3}$, 1034.2~$\rm pc\,cm^{-3}$, 1245~$\rm pc\,cm^{-3}$ and 1450.3~$\rm pc\,cm^{-3}$.
Note that this range is within the selected region demonstrated in \cite{Peng:2024aa}, which is regarded as reconstructions with relatively higher accuracy.
We derive the redshift distributions in these five broad DM bins by combining the reconstructed results from corresponding DM bins under the current thin binning scheme.
Since the large scatter between the $\rm DM_E$ and redshifts poses a significant challenge to the reconstruction of the self-calibration algorithm, we perform some additional operations to obtain higher accuracy.
Firstly, the reconstructed redshift distributions from the normalized median values of each scattering matrix element are shifted by an offset $\delta z$ to yield a best fit to the mean redshifts estimated using median values from the results of 100 perturbations, which are expected to be less biased on mean redshifts.
Meanwhile, since the reconstruction in the wide bin near the redshift edges might be less accurate, we ignore the values from this wide bin to achieve higher accuracy of mean redshifts.
The results are presented as the mean and the 1$\sigma$ uncertainty derived from 30 mock datasets.

The reconstructed redshift distributions and mean redshifts in each DM bin are shown in Figure~\ref{fig:nz_recon}.
It is obvious that this method both successfully reconstructs the characteristics of redshift distributions and accurately determines the mean redshifts of FRBs in DM space.
The biases in the mean redshifts (divided by (1+$z_{\rm true}$)) for mock FRB data in five DM bins are approximately $0.020\pm0.005, 0.009\pm0.006, 0.005\pm0.005, 0.012\pm0.004, 0.006\pm0.004$, respectively.
We show the reconstruction for all mock FRB data between 384.8 and 1450.3~$\rm pc\,cm^{-3}$ in Figure~\ref{fig:nz_recon_sample0}, and the mean redshift bias is constrained to $\sim\!0.001\pm0.003 (1+z_{\rm true})$.
To illustrate the generality and robustness of the self-calibration method, in Figure~\ref{fig:nz_recon_compare} we also present the reconstructions with the same detected FRB number, but the sky coverage during the one-year observation is increased from 1000~$\rm deg^2$ to 2000~$\rm deg^2$.
The bias in the mean redshift for all FRBs between 384.8 and 1450.3~$\rm pc\,cm^{-3}$ reaches $\sim\! 0.006\pm0.005 (1+z_{\rm true})$ due to higher noise level in this case.

\section{\label{sec:dis_cons}Discussion and Conclusions}
Determining the true redshift distribution of FRBs is critical for modeling sources and cosmological studies in future FRB surveys.
Since the localization precision of radio telescopes is typically poor and the host galaxies can be intrinsically faint, the direct spectroscopic observation of FRB sources is quite challenging and makes it difficult to obtain a large number of accurate redshift measurements.
We presented a novel method to address this issue by solely utilizing clustering in the DM space of observed FRBs to infer redshift distributions.
This approach does not require accurate modeling of the DM contributions from the host galaxy and the source's local environment.
We found that the method demonstrates robust and accurate performance in reconstructing the redshift distributions of mock FRB data, and the bias in the mean redshift can be constrained to within 1\%.
The performance of reconstructed results suggests that our method provides an effective solution for addressing the redshift issue in FRB observations with poor localization, making full use of the data to advance the study of cosmology and astrophysics.

Although these results are promising, taking into account the constraints of the current approach can provide significant insights for future studies and applications.
The accurate constraints on FRB redshift distributions that originate from clustering in DM space rely on low noise levels, especially in cases with large scatter, rather than on carefully selected photometric samples in galaxy surveys.
Therefore, we focus on FRB observations with relatively limited sky coverage to suppress noise.
The requirement for high signal-to-noise ratio measurements makes it challenging to apply the method to current or some upcoming surveys, such as CHIME \citep{CHIME/FRB-Collaboration:2024aa}, 2000-antenna Deep Synoptic Array \citep[DSA-2000;][]{Hallinan:2019aa}, Hydrogen Intensity and Real-time Analysis eXperiment \citep[HIRAX;][]{Newburgh:2016aa}, and Canadian Hydrogen Observatory and Radio-transient Detector \cite[CHORD;][]{Vanderlinde:2019aa}, due to their limited number of detections.
The main target of this work is to validate the effectiveness of the method while not accounting for potential systematics (e.g., cosmic magnification) in the clustering signal.
If there are many FRBs with very large $\rm DM_{host}$, the reconstruction performance will be under threat.
The practical redshift distribution and DM uncertainties of detected FRBs will also be influenced by their luminosity, which is not considered in this work and is expected to have minimal impact on the main conclusions.
Additionally, the galaxy bias of FRB host galaxies may be complex, which can affect the accuracy of the reconstruction.
More nuanced future studies are needed to explore these influencing factors.
With the rapid development of FRB observations, FRB clustering offers a clear path forward to address the redshift issue in cosmological and astrophysical studies.
Furthermore, other statistics, such as cross correlation with galaxy samples \citep{Li:2019aa,Rafiei-Ravandi:2020aa,Rafiei-Ravandi:2021aa}, can potentially improve the current self-calibration accuracy \citep{Peng:2024aa} and also deserves further exploration.

\begin{acknowledgments}
This work was supported by the National Key R\&D Program of China  (Nos. 2023YFA1607800, 2023YFA1607802), the National Natural Science Foundation of China (grant No. 12273020), the China Manned Space Project with Nos. CMS-CSST-2021-A02 and CMS-CSST-2021-A03, the “111” Project of the Ministry of Education under grant No. B20019, and the sponsorship from Yangyang Development Fund.
This work made use of the Gravity Supercomputer at the Department of Astronomy, Shanghai Jiao Tong University.
\end{acknowledgments}


\bibliography{sample7}{}
\bibliographystyle{aasjournal}



\end{document}